\documentclass[twocolumn,aps,pra,nofootinbib,superscriptaddress,amsmath,amssymb,floatfix,balance]{revtex4-2}
\usepackage{amsmath}
\usepackage{mathrsfs}
\usepackage{graphicx}
\usepackage[dvipsnames]{xcolor}
\usepackage{bm}
\usepackage[
	colorlinks=true,
	linkcolor=blue,
	urlcolor=blue,
	anchorcolor=blue,
	citecolor=blue,
	bookmarksnumbered]{hyperref}
\usepackage{txfonts}
\usepackage{balance}

\usepackage[T1]{fontenc}
\allowdisplaybreaks[4]

\begin{document}

\title{ Interference of cavity light by a single atom acting as a double slit }

\author{Yijia Zhou}
\affiliation{Graduate School of China Academy of Engineering Physics, Beijing 100193, China}

\author{Xinwei Li}
\affiliation{Graduate School of China Academy of Engineering Physics, Beijing 100193, China}

\author{Weibin Li}
\affiliation{School of Physics and Astronomy and Centre for the Mathematics and Theoretical Physics of Quantum Non-Equilibrium Systems, University of Nottingham, Nottingham, NG7 2RD, United Kingdom}

\author{Hao Zhang}
\email{hzhang@gscaep.ac.cn}
\affiliation{Graduate School of China Academy of Engineering Physics, Beijing 100193, China}

\begin{abstract}

	Young's double-slit interference experiment is central to quantum mechanics. While it has been demonstrated that an array of atoms can produce interference in light, it is a fundamental question to ask whether a single atom can act as a double slit when prepared in a superposition of two separate positions. Cohen-Tannoudji \textit{et al.} [\href{http://www.worldscientific.com/doi/abs/10.1142/9789814538190}{\textit{Proceedings of the Tenth International Conference on Laser Spectroscopy}}, edited by M. Ducloy, E. Giacobino, and G. Camy (World Scientific, Singapore, 1992), pp. 3–14] showed that the cross section of the light scattered by a single atom is independent of the spatial separation. In this work, however, we show that when a single atom tunneling in a double well is coupled to an optical ring cavity, the interference phenomena arise if the tunneling rate is comparable to the cavity linewidth. Being driven by an external laser in the dispersive regime, the field emitted by the atom into the cavity exhibits an interference pattern when varying the double-well spacing. Super-Poissonian bunched light can also be generated near the destructive interference. Furthermore, we show that the atomic flux of the coherent tunneling motion generates directional cavity emission, which oscillates for many cycles before the decoherence of the atomic motion and the decay of the cavity photons. Our work opens ways to manipulate photons with controllable external states of atoms for quantum information applications and use cavity light as nondestructive measurements for many-body states of atomic systems.

\end{abstract}
\date{\today}
\maketitle

\section{Introduction}

Light passing through a double slit can exhibit interference fringes. With advancements in atom trapping technologies, it has been theoretically studied ~\cite{Vogel1985, Skornia2001, Wiegner2011, Damanet2016} and experimentally demonstrated that an array of atoms can produce interference effects for light~\cite{Eichmann1993, Reimann2015, Reimann2015, Neuzner2016, Norcia2016, Kim2018, Wolf2020, Glicenstein2020, Rui2020} with potential applications in quantum information ~\cite{Facchinetti2016, Asenjo-Garcia2017, Guimond2019, Masson2020, Bekenstein2020, Ballantine2021, Ferioli2021, Fernandez-Fernandez2022}. One intriguing question to ask is whether a single atom, instead of an array of multiple atoms, can act as a double slit when a single atom is prepared in a linear superposition of two separated positions. Cohen-Tannoudji \textit{et al.}~\cite{Cohen-Tannoudji1992} gave a negative answer. The cross section of the light scattered by a single atom is independent of the atomic spatial coherence. There is no interference fringe when varying the separation of two superposed atomic wave packets, because the scattering process involves two orthogonal atomic spatial states.

Braun and Martin~\cite{Braun2008} studied the spontaneous emission of a single atom tunneling in a double well. They found that the emitted light can show interference fringes, but their amplitude is very small and on the order of $J/\omega_0$, where $J$ is the tunneling frequency and $\omega_0$ is the light frequency. In practice, $J$ can be many orders smaller than $\omega_0$ as $J/2\pi \sim 300~{\rm Hz}$~\cite{Kaufman2014} and $\omega_0/2\pi \sim 10^{14}~{\rm Hz}$. However, they proposed that if the external state of the atom is postselected in the energy basis or, alternatively, the light is spectrally filtered with a frequency resolution better than $J$, interference fringes with perfect visibility can arise.

\begin{figure}[b]
	\centering
	\includegraphics[width=\linewidth]{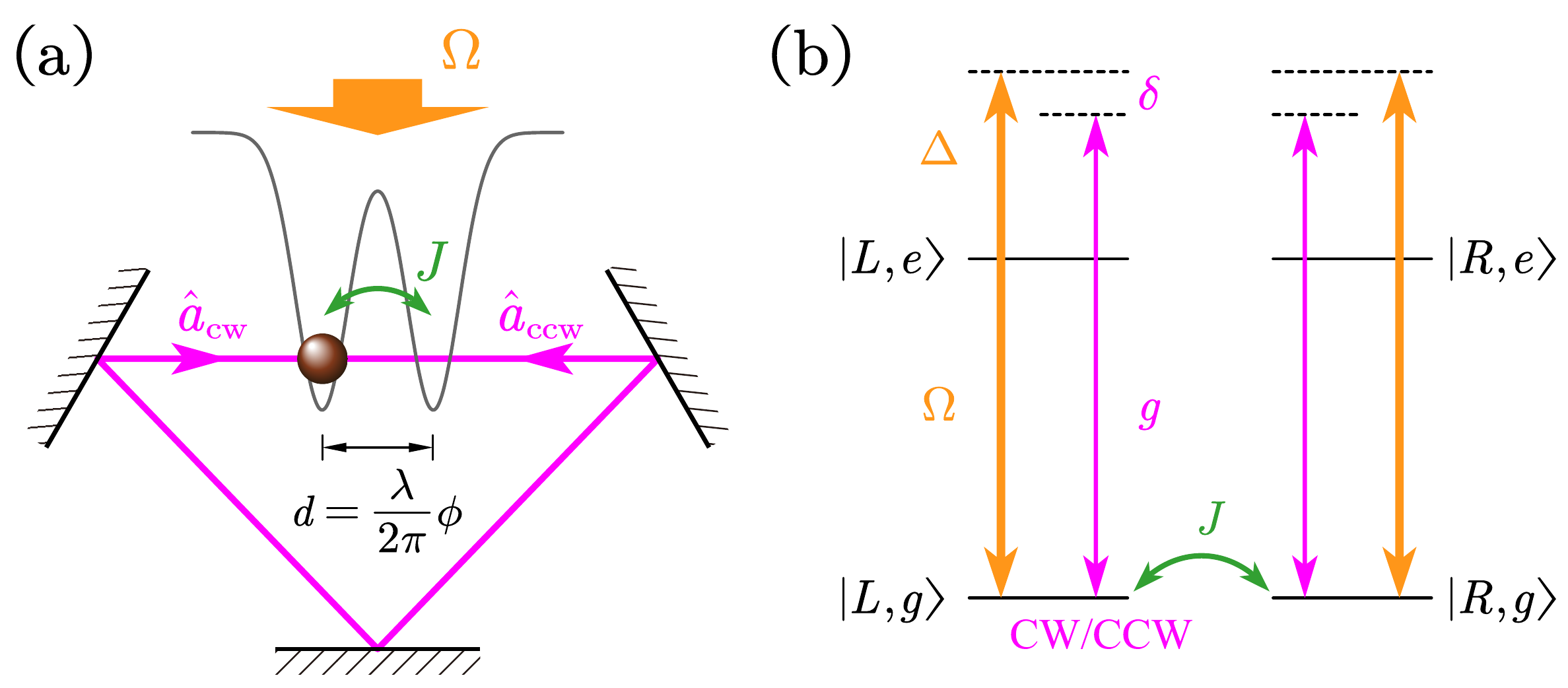}
	\caption{ \label{fig:Apparatus} (a) A single atom is confined in a double-well potential with spacing $d$ and tunneling amplitude $J$. The atom couples to an optical ring cavity with two counterpropagating modes $\hat{a}_{\rm CW}$ and $\hat{a}_{\rm CCW}$ and is driven by an external light field with the Rabi frequency $\Omega$.
	(b) The driving field is detuned from the atomic resonance by $\Delta$, and the cavity modes by $\delta$. The external states of the atom are denoted by $|L\rangle$ and $|R\rangle$.
	}
\end{figure}

In this paper we present an approach to manipulating and observing clear interference effects of light with a single atom acting as a double slit. A ring cavity is coupled to the atom, which is split in a double-well potential with the tunneling frequency $J$, as shown in Fig.~\ref{fig:Apparatus}(a). The atom is driven by an external laser and emits light into the cavity. Interference in the cavity emission can arise when varying the double-well spacing if $J$ is larger than the cavity linewidth $\kappa$. The role of the cavity is to store the light emitted by the atom at the left well for a time approximately equal to $1/\kappa$. If the cavity storage time is longer than the tunneling time, the atom quickly tunnels to the right well and emits light again such that the two light fields can interfere. The reason for using a ring cavity is that it supports traveling waves, allowing for the generation of an interference pattern when changing the double-well spacing.
In addition, the atomic motion significantly influences the photon correlations of the cavity emission and can yield super-Poissonian bunched light near the destructive interference. Furthermore, we show that the coherent tunneling motion of the single atom can be used to steer the light propagation direction, giving rise to the cavity directional emissions oscillating with the atomic flux in the double well. It also provides a non-destructive approach to monitoring atomic dynamics through cavity emission. 

\begin{figure} [t]
	\centering
	\includegraphics[width=\linewidth]{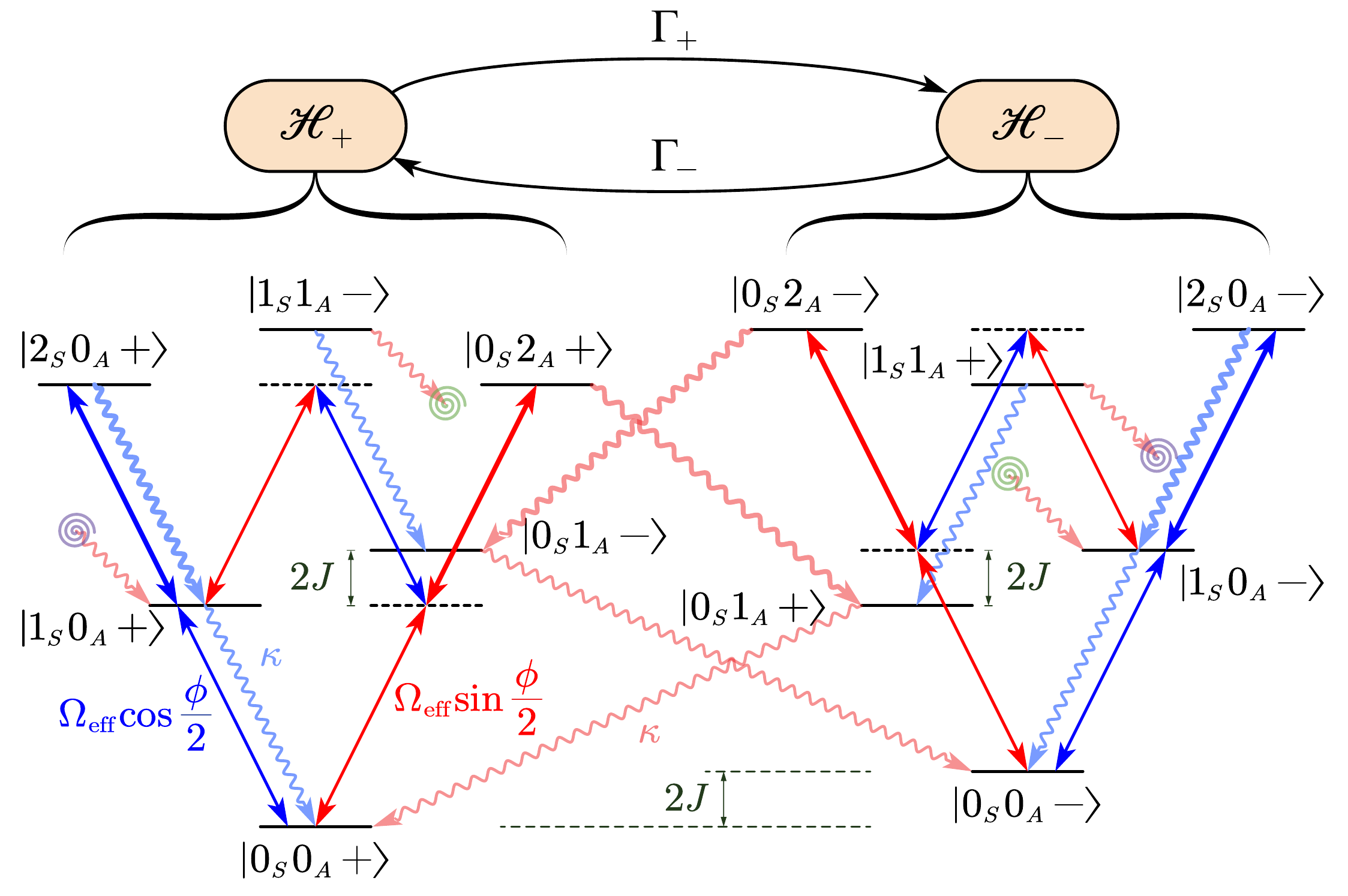}
	\caption{ \label{fig:Level_diagram} 
	Energy-level diagram revealing the $\mathbb{Z}_2$ symmetry after transforming the external states from the $\{|L\rangle, |R\rangle\}$ to the $\{|+\rangle, |-\rangle\}$ basis and the cavity fields from $\{ \hat{a}_{\rm CW}, \hat{a}_{\rm CCW} \}$ to $\{ \hat{a}_S, \hat{a}_A \}$ modes. The energy splitting between $|+\rangle$ and $|-\rangle$ is $2J$. The excited state $|e\rangle$ is eliminated in the dispersive regime. The $\mathscr{H_+}$ subspace is spanned by the even-parity states and $\mathscr{H_-}$ by the odd-parity states. The $\hat{a}_S$ mode is driven with the coupling strength $\Omega_{\rm eff}\cos\frac{\phi}{2}$ (blue arrows) and the $\hat{a}_A$ mode with $\Omega_{\rm eff}\sin\frac{\phi}{2}$ (red arrows). The cavity decays of the $\hat{a}_S$ and $\hat{a}_A$ modes are indicated by the blue and red wavy lines, respectively, at the rate $\kappa$ and do not change the external states $|\pm\rangle$.  However, the decay of the $\hat{a}_A$ mode photons leads to the population transfer between the subspaces $\mathscr{H}_+$ and $\mathscr{H}_-$ at the rates $\Gamma_+$ and $\Gamma_-$, which are proportional to the populations of $|0_S1_A-\rangle$ and $|0_S1_A+\rangle$ states, respectively. The purple and green spirals connect the relevant decay lines from the $|1_S1_A\pm\rangle$ states for illustration purposes. }
\end{figure}

\section{Model and Symmetry Analysis}

Our apparatus is shown in Fig.~\ref{fig:Apparatus}(a). We employ a ring-shaped optical cavity that supports two counterpropagating modes, denoted by the field operators $\hat{a}_{\rm CW}$ and $\hat{a}_{\rm CCW}$ for the clockwise (CW) and counterclockwise (CCW) directions, respectively. The atom is driven by an external laser field that couples two internal states $\left| g \right\rangle$ and $\left| e \right\rangle$ with Rabi frequency $\Omega$ as shown in Fig.~\ref{fig:Apparatus}(b). The detuning between the driving laser and the atomic resonance is $\Delta = \omega - \omega_a$ and the cavity detuning is $\delta = \omega - \omega_c$. The atomic tunneling amplitude between the two wells is $J$. The atom-photon interaction depends on not only the atom-cavity coupling $g$ but also a phase factor $e^{\pm i k z_j}$, where $z_j$ refers to the position of the atom along the cavity axis and $k=\omega_c/c$ is the wave number of the cavity modes. We have chosen to use a ring cavity because it supports two counterpropagating traveling waves with phase factors $e^{\pm i k z_j}$ allowing for the generation of interference in the cavity. The phases $e^{\pm i k z_j}$ are not present in Fabry-P\'erot cavities employed in studies of cavity coupling of atomic motion in double wells~\cite{Martin2008, Chen2009}. The external states of the atom are denoted by $|L\rangle$ and $|R\rangle$ centered at $z_L=-d/2$ and $z_R=d/2$ with $d$ the double-well spacing. The spatial phase difference is given by $\phi=2\pi d/\lambda$. The system is described by the Hamiltonian
\begin{equation} \label{eq:JCM}
	\begin{split}
		\hat{H} &= - \delta \left( \hat{a}_{\rm CW}^\dagger\hat{a}_{\rm CW} + \hat{a}_{\rm CCW}^\dagger\hat{a}_{\rm CCW} \right) - \Delta \hat{\sigma}^+\hat{\sigma}^- + \frac{\Omega}{2} \hat{\sigma}^x  \\
		&\phantom{{}=} + g \left[ \hat{\sigma}^+ \sum_{j=L, R} \left( e^{i \phi_j} \hat{a}_{\rm CW} + e^{-i \phi_j} \hat{a}_{\rm CCW} \right) |j\rangle\langle j| + {\rm H.c.} \right] \\
		&\phantom{{}=} - J \left( |L\rangle\langle R| + |R\rangle\langle L| \right) ,
	\end{split}
\end{equation}
where $\hat{\sigma}^\pm$ and $\hat{\sigma}^x$ are the Pauli operators associated with $|g\rangle$ and $|e\rangle$, and $\phi_L = -\phi/2$ and $\phi_R = \phi/2$. Including the atomic spontaneous emission rate $\gamma$ and the cavity decay rate $\kappa$, the evolution of the total density matrix follows the Lindblad master equation, 
\begin{equation} \label{eq:me}
	\dot{\rho} = - i [ \hat{H}, \rho ] + \gamma \mathcal{D}[\hat{\sigma}^-]\rho + \kappa \mathcal{D}[\hat{a}_{\rm CW}]\rho + \kappa \mathcal{D}[\hat{a}_{\rm CCW}]\rho ,
\end{equation}
where $\mathcal{D}[\hat{L}]\rho = \hat{L}\rho\hat{L}^\dagger - \frac12 \{ \hat{L}^\dagger\hat{L}, \rho \}$.

We notice that the Hamiltonian~\eqref{eq:JCM} has a $\mathbb{Z}_2$ symmetry, which can be better understood if we transform the atomic external state from the localized basis $\{ |L\rangle, |R\rangle \}$ to the extended basis $\{ |+\rangle=(|L\rangle+|R\rangle)/\sqrt{2}, |-\rangle=(|L\rangle-|R\rangle)/\sqrt{2} \}$ and the cavity fields from the $\hat{a}_{\rm CW}$ and $\hat{a}_{\rm CCW}$ modes to $\hat{a}_S = (\hat{a}_{\rm CW} + \hat{a}_{\rm CCW})/\sqrt{2}$ and $\hat{a}_A = -i(\hat{a}_{\rm CW} - \hat{a}_{\rm CCW})/\sqrt{2}$. The total photon number $n_{\rm tot}$ is conserved in both representations, such that $n_{\rm tot} = \langle \hat{a}_{\rm CW}^\dag \hat{a}_{\rm CW} \rangle + \langle \hat{a}_{\rm CCW}^\dag \hat{a}_{\rm CCW} \rangle = \langle \hat{a}_{S}^\dag \hat{a}_{S} \rangle + \langle \hat{a}_{A}^\dag \hat{a}_{A} \rangle $. In this paper we focus on the dispersive regime where $g,\Omega \ll |\Delta|$ and the excited state $|e\rangle$ is adiabatically eliminated (see Appendix~\ref{appx:sol}). The Hamiltonian of Eq.~\eqref{eq:JCM} is simplified to $H_{\rm eff} = H_S + H_A - J \hat{\sigma}_{\rm ext}^{z}$ with
\begin{subequations} \label{eq:H_SA}
	\begin{align} 
		H_S &= -\delta \hat{a}_S^\dagger \hat{a}_S + \frac{\Omega_{\rm eff}}{2} \cos\frac{\phi}{2} \left( \hat{a}_S^\dagger + \hat{a}_S \right) , \label{eq:H_S} \\
		H_A &= -\delta \hat{a}_A^\dagger \hat{a}_A + \frac{\Omega_{\rm eff}}{2} \sin\frac{\phi}{2} \left( \hat{a}_A^\dagger + \hat{a}_A \right) \hat{\sigma}_{\rm ext}^{x} , \label{eq:H_A} 
	\end{align}
\end{subequations}
where $\hat{\sigma}_{\rm ext}^{x} = |+\rangle\langle -| + |-\rangle\langle +|$, $\hat{\sigma}_{\rm ext}^{z} = |+\rangle\langle +| - |-\rangle\langle -|$, and $\Omega_{\rm eff}=\sqrt{2}g\Omega/\Delta$. The second term, $H_A$, is similar to the quantum Rabi model~\cite{Braak2011, Chen2012}. The $\mathbb{Z}_2$ symmetry of $H_{\rm eff}$ is revealed by the parity operator $\hat{\Pi} = \exp(i\pi \hat{a}_A^\dag \hat{a}_A) \hat{\sigma}_{\rm ext}^{z}$, which satisfies $[\hat{\Pi}, H_{\rm eff}] = 0$. Therefore, one can decompose the Hilbert space of $H_{\rm eff}$ into two subspaces according to the parity of the states, $\mathscr{H} = \mathscr{H}_+ \oplus \mathscr{H}_-$. Figure~\ref{fig:Level_diagram} shows the lowest 12 states categorized by the parity. For example, $|0_S,0_A,+\rangle \in \mathscr{H}_+$ and $|0_S,1_A,+\rangle \in \mathscr{H}_-$.
The $H_S$ couples the states $|n_S, n_A, \pm \rangle$ to $|n_S\pm1, n_A, \pm \rangle$ with the coupling strength $\Omega_{\rm eff}\cos\frac{\phi}{2}$ (the blue arrows in Fig.~\ref{fig:Level_diagram}), while $H_A$ couples the states $|n_S, n_A, \pm \rangle$ to $|n_S, n_A\pm1, \mp \rangle$ with the coupling strength $\Omega_{\rm eff}\sin\frac{\phi}{2}$ (the red arrows in Fig.~\ref{fig:Level_diagram}). While $H_{\rm eff}$ conserves the parity, the dissipation of the $\hat{a}_A$ mode photon leads to the incoherent population transfer between $\mathscr{H}_+$ and $\mathscr{H}_-$ with the respective rates $\Gamma_+$ and $\Gamma_-$ calculated below.

\begin{figure}
	\centering
	\includegraphics[width=0.8\linewidth]{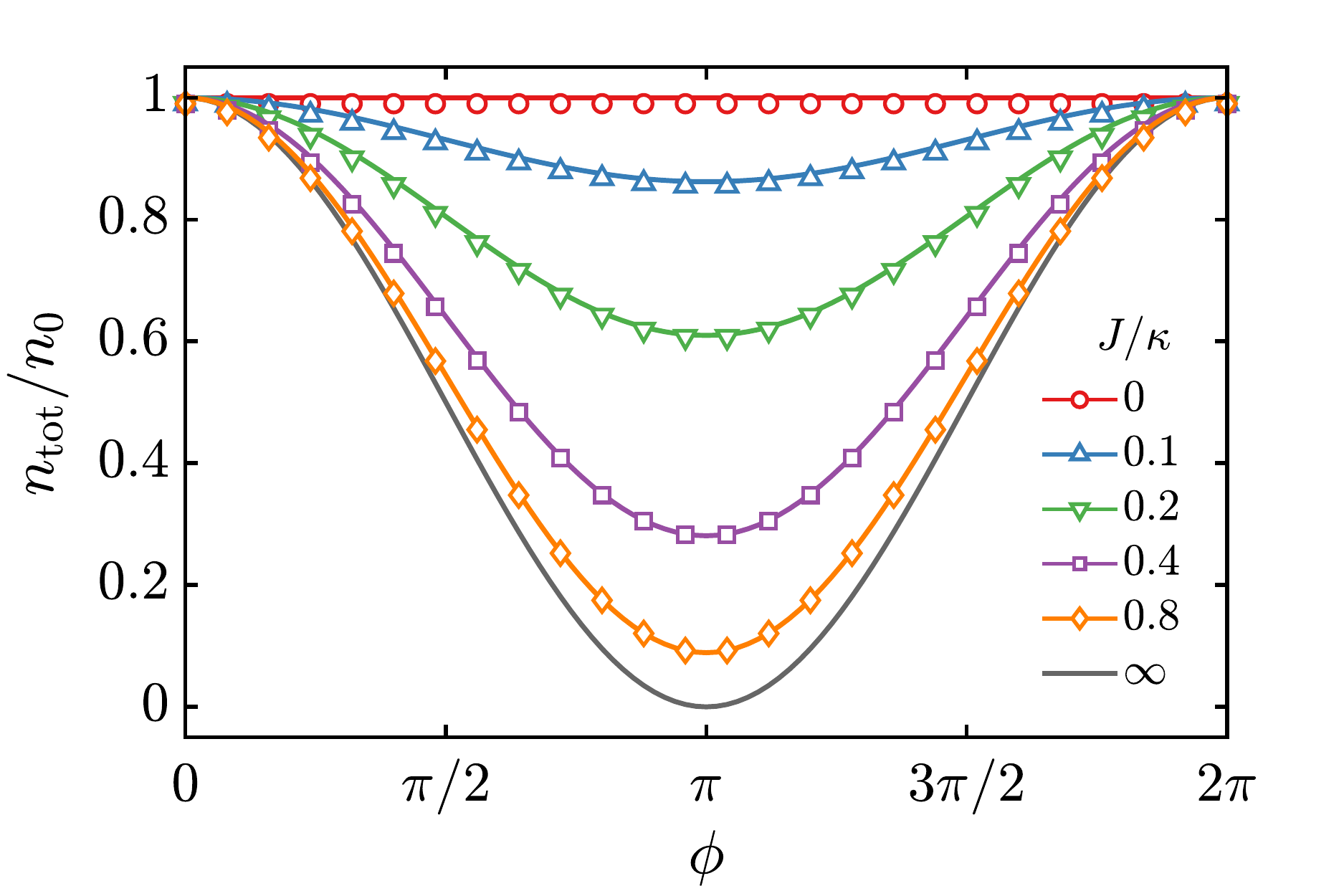}
	\caption{ \label{fig:steady_solution} Interference of the cavity field when varying the double-well spacing. The single atom is driven by the external laser on the cavity resonance ($\delta=0$).
	The steady-state cavity photon number $n_{\rm tot}$ shows a sinusoidal interference pattern as a function of the double-well spacing denoted by $\phi=2\pi d/\lambda$, where $n_{\rm tot}$ is normalized by the photon number $n_{0}$ of an atom trapped in a fixed position. The interference contrast increases with $J/\kappa$. The solid lines are the analytical results according to Eq.~\eqref{eq:steady} and the data points are the numerical solutions to the master equation \eqref{eq:me}. The other parameters are $(\gamma, \Delta, g, \Omega)/\kappa = (10, 200, 0.5, 20)$, which are also applied throughout this work. }
\end{figure}

\section{Interference of the Cavity Light}

Using $H_{\rm eff}$ expressed in the $\hat{a}_S$ and $\hat{a}_A$ basis [Eq.~\eqref{eq:H_SA}], we calculate the total cavity photon number $n_{\rm tot}$ of the steady state (see Appendices~\ref{appx:sol} and \ref{appx:dyn}) as
\begin{equation} \label{eq:steady}
	n_{\rm tot} \simeq n_{0} \left( \cos^2\frac{\phi}{2} +  \frac{1}{1 + 4J^2/(\delta^2+\kappa^2/4)} \sin^2\frac{\phi}{2} \right) ,
\end{equation}
where $n_{0} = 2 \times \left|(g\Omega/2\Delta)/(\delta+i\kappa/2) \right|^2 $ is for an atom in a fixed position.
The first term in Eq.~\eqref{eq:steady} is the photon number of the $\hat{a}_S$ mode driven by $|0_S 0_A \pm\rangle \leftrightarrow |1_S 0_A \pm\rangle$ with the transition amplitude $\Omega_{\rm eff}\cos\frac{\phi}{2}$. This term is in fact unrelated to the atomic external state. The second term is the $\hat{a}_A$ mode driven by $|0_S 0_A \pm\rangle \leftrightarrow |0_S 1_A \mp\rangle$ with the transition amplitude $\Omega_{\rm eff}\sin\frac{\phi}{2}$ and the detuning $\pm2J$. Unlike the $\hat{a}_S$ mode, the photon number of the $\hat{a}_A$ mode is related to the final population in the subspaces $\mathscr{H}_+$ and $\mathscr{H}_-$, which is determined by the tunneling rate $J$ and the detuning $\delta$. The detailed derivation also includes the decoherence process, which is discussed in Appendix~\ref{appx:dyn}.

Figure~\ref{fig:steady_solution} shows the sinusoidal modulation of the cavity photon number $n_{\rm tot}$ as a function of the double-well spacing on the cavity resonance ($\delta=0$). The interference contrast of $n_{\rm tot}$ increases with $J$. When $J=0$, the photon number is independent of $\phi$ ($n_{\rm tot} = n_0$), which is consistent with Refs.~\cite{Cohen-Tannoudji1992, Braun2008}. Note that as long as $J=0$, regardless of whether the atom is in a superposition state of $|L\rangle$ and $|R\rangle$ or a completely mixed state, the cavity field does not exhibit interference. However, in the limit $J/\kappa \to \infty$, the interference with perfect contrast appears when the second term of Eq.~\eqref{eq:steady} vanishes and $n_{\rm tot} \to n_0 \cos^2\frac{\phi}{2}$. This can be understood by the level diagram of Fig.~\ref{fig:Level_diagram}. The transitions generating the $\hat{a}_S$ mode photons from $|0_S, 0_A,\pm\rangle$ to $|1_S, 0_A,\pm\rangle$ (the blue arrows in Fig.~\ref{fig:Level_diagram}) are always on-resonance since $\delta = 0$, while the transitions generating the $\hat{a}_A$ mode photons from $|0_S, 0_A,\pm\rangle$ to $|0_S, 1_A,\mp\rangle$ (the red arrows in Fig.~\ref{fig:Level_diagram}) are detuned by $\pm 2J$. When $J > \kappa$, the generation of the $\hat{a}_A$ mode photons is suppressed, so the cavity field is dominated by the $\hat{a}_S$ mode, resulting in the interference pattern which is proportional to $\cos^2\frac{\phi}{2}$. From a complementary perspective, the light emitted by the atom at the position $|L\rangle$ is stored in the cavity over time approximately equal to $1/\kappa$. After the tunneling time approximately equal to $1/J$, the atom emits light from the position $|R\rangle$. The two light fields can interfere if the tunneling time is shorter than the cavity decay time ($1/J < 1/\kappa$).

\section{Cavity Photon Correlations}

\begin{figure}
	\includegraphics[width=\linewidth]{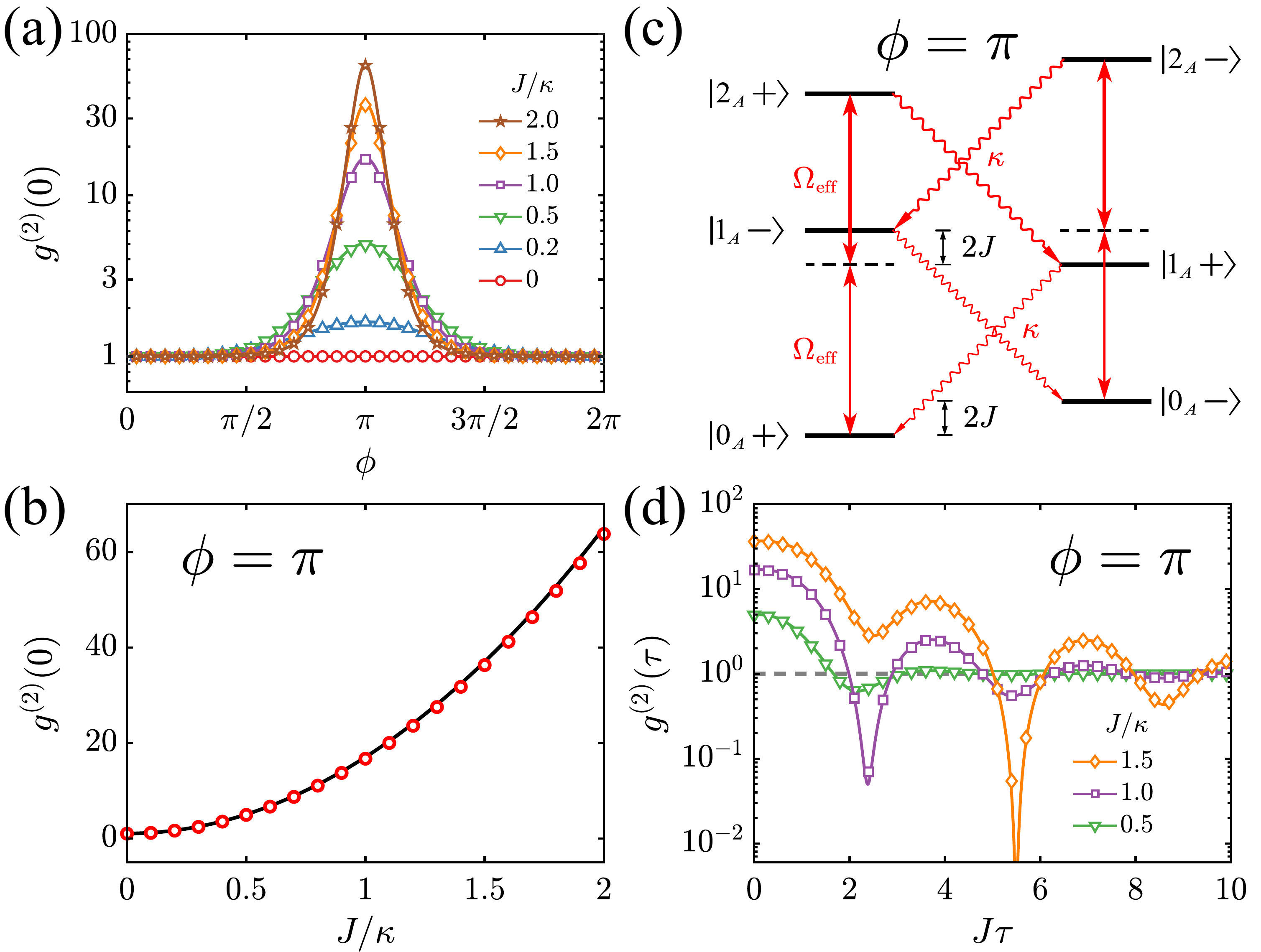}
	\caption{\label{fig:corr}  Second-order correlation functions of the cavity emission at $\delta=0$. (a) Plot of $g^{(2)}(0)$ as a function of $\phi$ for different values of $J/\kappa$, indicating transitions from Poissonian near $\phi=0$ to super-Poissonian near $\phi=\pi$. (b) Maximum values of $g^{(2)}(0)$ at $\phi=\pi$ compared with $g^{(2)}(0) = 1+16J^2/\kappa^2$ (black solid line). (c) Level diagram for $\phi=\pi$. The $\hat{a}_S$ mode photons in Fig.~\ref{fig:Level_diagram} are not excited due to the destructive interference and hence are not shown. The single-photon excitations of the $\hat{a}_A$ mode are detuned by $\pm2J$, which leads to the photon-pair generation and the photon bunching effect. (d) The $g^{(2)}(\tau)$ at $\phi=\pi$ for different values of $J/\kappa$ shows stronger oscillation amplitudes for larger tunneling $J$. The oscillation frequency is $2J$ and the damping rate is given by $\kappa$. In (a), (b) and (d) the solid lines represent the analytical solutions of Eq.~\eqref{eq:g2} and the data points correspond to the numerical simulation results. } 
\end{figure}

The interference induced by the atomic tunneling has a strong effect on the cavity photon statistics, which can be characterized by the second-order correlation function $g^{(2)}(\tau) = \langle \hat{a}^\dag \hat{a}^\dag(\tau) \hat{a}(\tau) \hat{a} \rangle / \langle \hat{a}^\dag \hat{a} \rangle^2$, with $\hat{a}$ either $\hat{a}_{\rm CW}$ or $\hat{a}_{\rm CCW}$. The denominator contains the photon number  $\langle \hat{a}^\dag \hat{a} \rangle$ of the steady state $\rho_{\rm ss}$. The numerator is calculated in the Schr\"odinger picture~\cite{Goncalves2021} as $\langle \hat{a}^\dag \hat{a}^\dag(\tau) \hat{a}(\tau) \hat{a} \rangle = {\rm Tr} [ \hat{a}^\dag \hat{a} \rho'(\tau) ] {\rm Tr} [\hat{a}^\dag \hat{a} \rho_{\rm ss} ]$, where  $\rho'(\tau)$ evolves from the conditional state by projection $\rho'(0) = \hat{a} \rho_{\rm ss} \hat{a}^\dag / {\rm Tr} [\hat{a}^\dag \hat{a} \rho_{\rm ss} ]$ (see Appendix~\ref{appx:corr} for details). Finally, we find that the analytical solution of $g^{(2)}(\tau)$ is the same for the two modes and reads 
\begin{equation} \label{eq:g2}
	\begin{split}
		g^{(2)}(\tau) &= 1 + \left[ 1 + \left( 1 + \frac{16J^2}{\kappa^2} \right) \cot^2\frac{\phi}{2} \right]^{-2} \\
		& \qquad \times \left( \frac{16J^2}{\kappa^2} e^{-\kappa \tau} + \frac{8J}{\kappa} e^{-\kappa \tau / 2} \sin 2J\tau 
			\right) .
	\end{split}
\end{equation}
The equal-time correlation $g^{(2)}(0)$ of the cavity emission is plotted in Fig.~\ref{fig:corr}(a) as a function of $\phi$, showing strong photon bunching near the destructive interference $\phi=\pi$. The peak value of $g^{(2)}(0)$ at $\phi=\pi$ increases quadratically with $J/\kappa$, as shown in Fig.~\ref{fig:corr}(b).
At the destructive interference, the transition amplitudes of the $\hat{a}_S$ mode (the blue arrows in Fig.~\ref{fig:Level_diagram}) are canceled out as $\Omega_{\rm eff}\cos\frac{\phi}{2}=0$, and only the $\hat{a}_A$ mode photons are generated as $\Omega_{\rm eff}\sin\frac{\phi}{2}=\Omega_{\rm eff}$. The level diagram of Fig.~\ref{fig:Level_diagram} is simplified to Fig.~\ref{fig:corr}(c).
When $J > \kappa$, the one-photon excitations ($|0_A\rangle \to |1_A\rangle$) are far-off-resonance and hence suppressed, while the simultaneous excitations of two photons ($|0_A\rangle \to |1_A\rangle \to |2_A\rangle$) are on-resonance, resulting in the photon-pair generation and huge photon bunching. 
The correlation $g^{(2)}(\tau)$ displays a temporal oscillation with the frequency $2J$ due to the cavity detuning from the $|0_A \pm \rangle \to |1_A \mp \rangle$ transitions as shown in Fig.~\ref{fig:corr}(d).

\section{Directional Cavity Emission Induced by Atomic Motion}

\begin{figure} [!h]
	\centering
	\includegraphics[width=\linewidth]{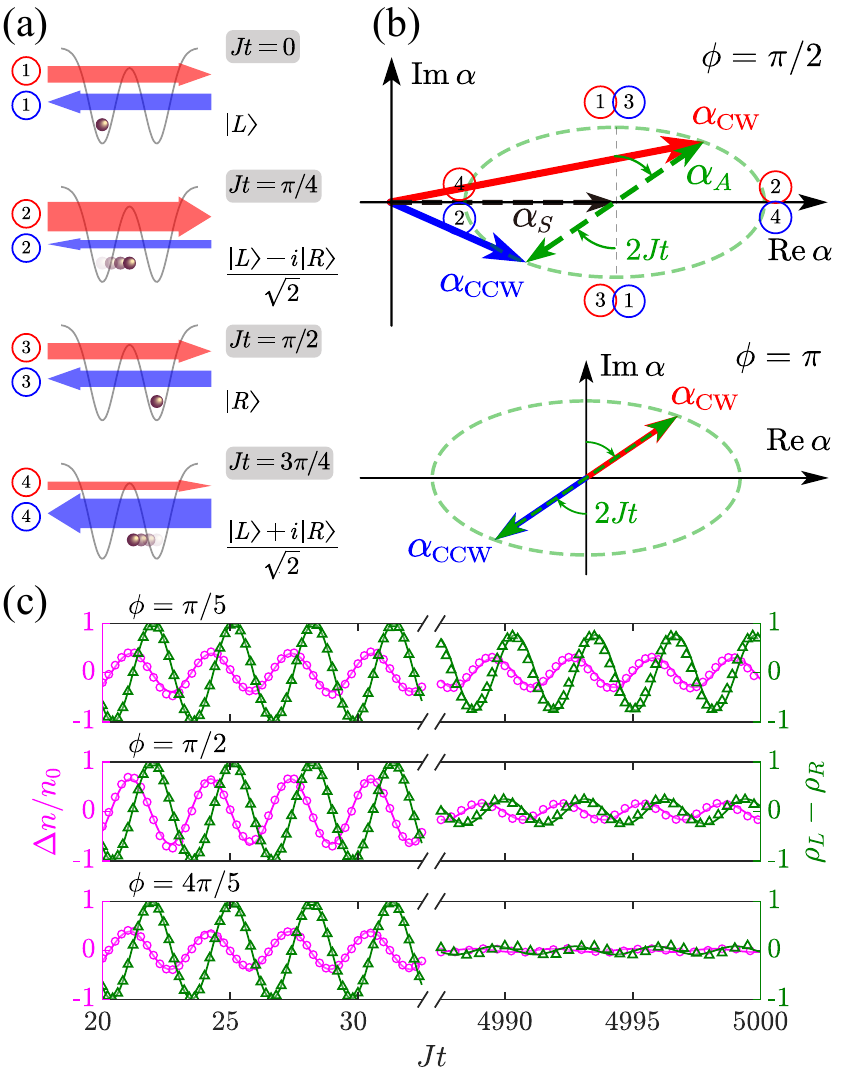}
	\caption{ \label{fig:chirality} 
	Cavity light directionality induced by the atomic motion.
	(a) The CW (red arrow) and CCW (blue arrow) modes are shown at different times of the atomic tunneling motion for $\phi=\pi/2$. When the atom is in the superposition states $\frac{1}{\sqrt{2}} (|L\rangle \pm i |R\rangle)$ with the maximum tunneling flux, the cavity fields have different intensities for the CW and CCW modes.
	(b) Sketch of the instantaneous cavity fields in the phase space for $\phi=\pi/2$ and $\pi$, where $\alpha=\langle\hat{a}\rangle$. The CW (red) and CCW (blue) modes are constructed by the $\hat{a}_S$ (black) and $\hat{a}_A$ (green) modes according to $\alpha_{\rm CW,CCW} = (\alpha_{S} \pm i \alpha_A)/\sqrt{2}$. The $\hat{a}_S$ mode is time independent, while the $\hat{a}_A$ mode rotates at a frequency of $2J$. The green dashed ellipses represent the trajectories of $\alpha_{\rm CW}$ and $\alpha_{\rm CCW}$. When $\phi=\pi/2$, the magnitudes of $\alpha_{\rm CW}$ and $\alpha_{\rm CCW}$ oscillate at a frequency of $2J$. When $\phi=\pi$, $\alpha_S = 0$ and therefore $|\alpha_{\rm CW}| = |\alpha_{\rm CCW}|$.
	The circled numbers correspond to the time steps in (a).
	(c) Numerical simulations of the dynamical evolution of $\Delta{n}/n_0 = (n_{\rm CW} - n_{\rm CCW})/n_0$ (magenta circles) and the atomic population difference between the left and right wells $\rho_{L} - \rho_{R}$ (green triangles). Solid lines represent analytical results. Both the cavity light directionality $\Delta{n}/n_0$ and the atomic position $\rho_L - \rho_R$ oscillate at a frequency approximately equal to $2J$. Their phase difference is $\pi/2$. When the double-well spacing increases from $\phi=\pi/5$ to $\phi=4\pi/5$, the decoherence rate $\Gamma$ of the atomic tunneling increases due to the backaction of the cavity light, resulting in the damping of the directional cavity emissions. The initial state of the atom is $|L\rangle$ with $-\delta = J = 5 \kappa$.
	}
\end{figure}

Having studied the steady-state cavity interference and the photon correlations, we next examine the dynamical evolution of the cavity field before reaching equilibrium. As the atom tunnels within the double well, directional emissions from the cavity can be generated, which oscillate at the tunneling frequency $J$ as depicted in Fig.~\ref{fig:chirality}(a). On the other hand, the backaction of the cavity field gives rise to the decoherence of the atomic tunneling motion. If the tunneling frequency $J$ is large compared to the decoherence rate $\Gamma$ (which will be derived later), the instantaneous cavity fields adiabatically follow the external state of the atom described by the reduced density matrix $\rho_{\rm ext}(t)$. To understand the physical picture, we examine the cavity fields in the mean-field approximation as $\alpha_\mu = \langle\hat{a}_\mu\rangle$, where $\hat{a}_\mu$ is either the CW or CCW mode or the $\hat{a}_S$ or $\hat{a}_A$ mode. In the adiabatic approximation [see Eq.~\eqref{eq:mf_alpha} in Appendix~\ref{appx:sol}]
\begin{subequations} \label{eq:light_class}
	\begin{align}
		\alpha_S &= \frac{\Omega_{\rm eff}}{2} \frac{2}{\delta + i\kappa/2} \cos\frac{\phi}{2} ,\\
		\alpha_A &= \frac{\Omega_{\rm eff}}{2} \left(
			\frac{\rho_{\rm ext}^{+-}(t)}{\delta -2J + i\kappa/2} +
			\frac{\rho_{\rm ext}^{-+}(t)}{\delta + 2J + i\kappa/2} \right) \sin\frac{\phi}{2}  .
	\end{align}
\end{subequations}
At the early time $t \ll 1/\Gamma$ before the tunneling motion decoheres, $\rho_{\rm ext}(t) \approx \frac{1}{2} ( |+\rangle\langle+| + |-\rangle\langle-| + e^{2iJt} |+\rangle\langle-| + e^{-2iJt} |-\rangle\langle+| )$. Substituting $\rho_{\rm ext}(t)$ into Eq.~\eqref{eq:light_class}, the evolution of $\alpha_{\rm CW,CCW}$ traces an ellipse in the phase space as shown in Fig.~\ref{fig:chirality}(b). When $\phi=\pi/2$ and $\delta=-J$, the amplitude $|\alpha_S|$ is comparable to $|\alpha_A|$; therefore $|\alpha_{\rm CW}|$ can be very different from $|\alpha_{\rm CCW}|$, resulting in directional emissions. When $\phi=\pi$, however, $\alpha_S=0$; therefore the two fields have the same amplitudes but opposite phases, $\alpha_{\rm CW} = - \alpha_{\rm CCW}$. We define the cavity light directionality as the photon number difference between the two cavity modes $\hat{a}_{\rm CW}$ and $\hat{a}_{\rm CCW}$, 
\begin{equation} \label{eq:chirality}
	\Delta{n}/n_0 = (n_{\rm CW} - n_{\rm CCW})/n_0 ,
\end{equation}
where $n_{\rm CW,CCW} = \frac{1}{2} \langle (\hat{a}_S \pm i \hat{a}_A )^\dag (\hat{a}_S \pm i \hat{a}_A ) \rangle = \frac{1}{2}[n_S + n_A(t)] \pm {\rm Im}[ \alpha_S  \alpha_A^*(t) ]$ [see Eq.~\eqref{eq:mf_n}]. 
From Eqs.~\eqref{eq:light_class} and \eqref{eq:chirality}, $\Delta{n}/n_0 \propto \sin\phi \ {\rm Im}\rho_{\rm ext}^{+-}$ when $\kappa \ll |\delta \pm 2J|$. Notice that ${\rm Im}\rho_{\rm ext}^{+-} = {\rm Im} \langle L|\rho_{\rm ext}|R\rangle $ indicates the atomic flux of the tunneling motion from $|L\rangle$ to $|R\rangle$ in the double well. In Fig.~\ref{fig:chirality}(a), at $Jt=0$ and $\pi/2$, the atomic flux is zero hence both the CW and CCW modes have the same photon numbers. At $Jt=\pi/4$ and $3\pi/4$, the atomic flux reaches the maximum value resulting in the cavity directional emission. Figure~\ref{fig:chirality}(c) shows the dynamical evolution of $\Delta{n}/n_0$ compared with the population difference in the double well $\rho_L-\rho_R$ for different values of $\phi$.

The backaction of the cavity field gives rise to the decoherence and frequency shift for the atomic tunneling motion. To calculate the decoherence rate $\Gamma$, we obtain the master equation for the atomic external state (see Appendix~\ref{appx:dyn})
\begin{equation} \label{eq:me_ext}
	\begin{split}		
	\dot{\rho}_{\rm ext} &\approx -i\left[-J\hat{\sigma}^{z}_{\rm ext} + J_+ \left| + \middle\rangle\middle\langle + \right| + J_- \left| - \middle\rangle\middle\langle - \right|,\ \rho_{\rm ext} \right] \\
	&\phantom{{}=} +  \Gamma_+ \mathcal{D}[\hat{\sigma}_{\rm ext}^+] \rho_{\rm ext} + \Gamma_- \mathcal{D}[\hat{\sigma}_{\rm ext}^-] \rho_{\rm ext} . 
	\end{split}
\end{equation}
The dissipator $\mathcal{D}[\hat{\sigma}_{\rm ext}^+]$ describes the incoherent population transfer from the subspace $\mathscr{H}_+$ to $\mathscr{H}_-$, and similarly $\mathcal{D}[\hat{\sigma}_{\rm ext}^-]$ the transfer from $\mathscr{H}_-$ to $\mathscr{H}_+$, as shown in Fig.~\ref{fig:Level_diagram}. The respective coefficients are given by 
\begin{equation} \label{eq:shift}
	J_\pm - i \frac{\Gamma_\pm}{2} = \frac{\Omega_{\rm eff}^2}{4} \frac{1}{\delta \mp 2J + i \kappa/2} \sin^2\frac{\phi}{2} .
\end{equation}
In Eq.~\eqref{eq:shift} the imaginary parts $\Gamma_\pm$ correspond to the cavity photon generation rate as $\kappa n_A(t) = \Gamma_+ \rho_{\rm ext}^{++}(t) + \Gamma_- \rho_{\rm ext}^{--}(t)$. The real parts $J_\pm$ change the original tunneling rate $J$ to $J' = J - (J_+ - J_-)/2$, resulting from the ac Stark shift of the cavity light. 
With the initial state being $\rho_{\rm ext}(0) = |L\rangle\langle L|$, the reduced density matrix of the external state evolves as $\rho_{\rm ext}^{-+} = e^{-\Gamma t/2} e^{i2J't}$, where $\Gamma=\Gamma_+ + \Gamma_-$ is modulated by $\sin^2\frac{\phi}{2}$.
As shown in Fig.~\ref{fig:chirality}(c), the decoherence is faster for $\phi\approx \pi$ than for $\phi\approx 0$.
In the dispersive regime, $\Gamma$ is several orders smaller than $\kappa$ and $J$, which allows for many cycles of oscillation before the decoherence of the atomic motion and the leakage of the cavity photons.

\section{Conclusions and Discussions}

In conclusion, we have demonstrated that a single atom can function as a double slit and produce pronounced interference effects for light. Specifically, when the atomic tunneling $J$ in the double well is larger than the ring cavity linewidth $\kappa$, the interference in the cavity light with perfect contrast can arise. Furthermore, the atom's coherent tunneling motion gives rise to photon-pair generation and cavity directional emissions. In practice, the tunneling rate $J$ between the two wells is on the order of $2\pi\times300~{\rm Hz}$, while the cavity linewidth $\kappa \sim 2\pi\times 5~{\rm kHz}$ can be realized experimentally~\cite{Schuster2020, Kedar2023}. In order to satisfy the condition $J> \kappa$, we propose to make use of higher excited external states that are more spatially extended in the trapping potential. By driving the $|L\rangle$ and $|R\rangle$ states to a higher extended state, the coupling $J$ between $|L\rangle$ and $|R\rangle$ increases with the driving power and can be made larger than $\kappa$. Therefore, our scheme of interfering light using a single atom as a double slit is practically achievable. 

It is worth noting that while we study the interference effect of light scattered by a single atom, previous studies have shown that interference can arise when a massive particle is scattered by a quantum obstacle that can coherently tunnel between two positions~\cite{Schomerus2002} or by another particle with a similar mass~\cite{Rohrlich2006}. In our work and Refs.~\cite{Schomerus2002, Rohrlich2006}, the double slit, being either a single atom in a double well or a quantum obstacle, is an active quantum object that can move or tunnel to different positions, as opposed to Young's double slit, which is stationary and only plays a passive role. When a massive particle is scattered, interference appears when the tunneling $J$ is greater than the particle's kinetic energy. In this paper, however, when a massless cavity photon is scattered by a massive atom, the condition for interference is $J>\kappa$.

This work can be readily extended in various directions. Both the atomic motional states in the double well and the directional cavity photons can be utilized to encode qubits. Their entanglement allows for new gate operation schemes and our apparatus may serve as quantum nodes in cavity-based quantum networks. Recent experiments of controllable directional photon emission and scattering in waveguide QED systems~\cite{Kannan2023, Redchenko2023} are based on two-qubit interference. Our theory shows the possibility of tuning photon propagating directions using only one particle. We can also extend our current minimal model of a single atom in a double well to many-atom systems in lattices. Cavity emissions can provide non-destructive measurements of atomic collective motion and many-body states, such as Bloch oscillations, superfluid-Mott insulator transitions, and self-organization.

\begin{acknowledgements}

	The authors thank Chang-Pu Sun and Markus M\"uller for insightful discussions. This work was supported by the National Key Research and Development Program of China (Grant  No.~2022YFA1405300) and the National Natural Science Foundation of China (Grant No.~12088101). WL acknowledges support from the International Research Collaboration Fund of the University of Nottingham.
	
\end{acknowledgements}

\appendix

\section{Analytical solutions of the photon fields in the dispersive regime}
\label{appx:sol}

The original Hamiltonian of the cavity QED system with a single atom coupled with two propagating cavity modes and trapped in a double-well potential under the rotating frame is
\begin{equation} \label{eq:original_H}
	\begin{split}
		H &= - \delta \left(\hat{a}_{\rm CW}^\dag\hat{a}_{\rm CW} + \hat{a}_{\rm CCW}^\dag\hat{a}_{\rm CCW}\right)
		- \Delta \left| e\middle\rangle \middle\langle e\right| \\
		&\phantom{{}=} + \frac{\Omega}{2} \hat{\sigma}^x - J \left( |L \rangle\langle R| + |R \rangle\langle L| \right) \\
		&\phantom{{}=} + g \left\{ \left[ \left( e^{i\mathbf{k}_{\rm CW}\cdot\mathbf{r}_L} \hat{a}_{\rm CW} + e^{i\mathbf{k}_{\rm CCW}\cdot\mathbf{r}_L} \hat{a}_{\rm CCW} \right) | L\rangle\langle L| + \right.\right.\\
		&\phantom{{}=} \left.\left. \left( e^{i\mathbf{k}_{\rm CW}\cdot\mathbf{r}_R} \hat{a}_{\rm CW} + e^{i\mathbf{k}_{\rm CCW}\cdot\mathbf{r}_R} \hat{a}_{\rm CCW} \right) | R\rangle\langle R| \right]  \hat{\sigma}^+ + {\rm H.c.} \right\},
	\end{split}
\end{equation}
where the external state $|L\rangle$ or $|R\rangle$ denotes the atomic position, which is along the $z$ axis, such that $\mathbf{r}_{L,R} = (0,0,z_{L,R})$. The wave vectors of the CW and CCW modes are $\mathbf{k}_{\rm CW} = (0,0,k)$ and $\mathbf{k}_{\rm CCW}=(0,0,-k)$ near the double well, respectively. Due to translational symmetry, it is convenient to define $k z_L = -\phi/2$ and $k z_R = \phi/2$, and the double-well spacing is defined as $d=(\phi/2\pi)\lambda$, with $\lambda$ the wavelength of the cavity fields. The atomic decay and photon loss can be described by the Lindblad master equation [Eq.~\eqref{eq:me}]
\begin{equation} \label{eq:original_ME}
	\frac{\partial}{\partial t}\rho = -i [H, \rho] + \gamma \mathcal{D}[\hat{\sigma}^-]\rho + \kappa \mathcal{D}[\hat{a}_{\rm CW}]\rho + \kappa \mathcal{D}[\hat{a}_{\rm CCW}]\rho ,
\end{equation}
where $\mathcal{D}[\hat{L}]\rho = \hat{L}\rho\hat{L}^\dag - \frac{1}{2} \{ \hat{L}^\dag\hat{L}, \rho \}$. Our dynamical simulations are based on the time integration of Eq.~\eqref{eq:original_ME} and the steady state by iteration.

The dispersive regime is valid when the atomic detuning is far bigger than the atomic and photonic dissipation rates and the atom-light interaction strength $\Delta \gg \gamma,\kappa,g$ such that the atomic internal state can reach equilibrium faster than other time scales. In this situation, the original Hamiltonian can be simplified by adiabatic elimination. 
First of all, it is convenient to use the superposed photon modes as described in the main text,
\begin{equation}
	\begin{split}
		\hat{a}_S &= \frac{1}{\sqrt{2}} \left( \hat{a}_{\rm CW} + \hat{a}_{\rm CCW} \right) ,\\
		\hat{a}_A &= \frac{-i}{\sqrt{2}} \left( \hat{a}_{\rm CW} - \hat{a}_{\rm CCW} \right) ,
	\end{split}
\end{equation}
which automatically guarantees that $\mathcal{D}[\hat{a}_{\rm CW}] + \mathcal{D}[\hat{a}_{\rm CCW}] = \mathcal{D}[\hat{a}_{S}] + \mathcal{D}[\hat{a}_{A}]$. The atomic external states should also be symmetrized as
\begin{equation}
	\begin{split}
		| + \rangle &= \frac{1}{\sqrt{2}} \left( | L \rangle + | R \rangle \right) ,\\
		| - \rangle &= \frac{1}{\sqrt{2}} \left( | L \rangle - | R \rangle \right) .
	\end{split}
\end{equation}
Then the tunneling term reads $H_{\rm ext} = -J \hat{\sigma}_{\rm ext}^z$, with the Pauli matrices for the external states being $\hat{\sigma}_{\rm ext}^z = |+\rangle\langle+| - |-\rangle\langle-|$ and $\hat{\sigma}_{\rm ext}^x = |+\rangle\langle-| + |-\rangle\langle+|$.
Then we rewrite the original master equation [Eq.~\eqref{eq:me} or \eqref{eq:original_ME}] as
\begin{equation}
	\begin{split}
	\frac{\partial}{\partial t}\rho &= -i \left[ \left( H_{g, \rm nH} + H_{e, \rm nH} \right) \rho - \rho \left( H_{g, \rm nH} + H_{e, \rm nH} \right)^\dag \right] \\
	&\phantom{{}=} -i \left[V + V^\dag, \rho \right] + \gamma \hat{\sigma}^-\rho\hat{\sigma}^+ \\
	&\phantom{{}=} + \kappa \hat{a}_{\rm CW}\rho\hat{a}_{\rm CW}^\dag + \kappa \hat{a}_{\rm CCW}\rho\hat{a}_{\rm CCW}^\dag ,
	\end{split}
\end{equation}
where
\begin{subequations}
\begin{align*}
		H_{g, \rm nH} &= - \left( \delta + i\frac{\kappa}{2} \right) \left( \hat{a}_S^\dag\hat{a}_S + \hat{a}_A^\dag\hat{a}_A \right) - J \hat{\sigma}_{\rm ext}^z ,\\
		H_{e, \rm nH} &= - \left( \Delta + i\frac{\gamma}{2} \right) \left| e\middle\rangle\middle\langle e\right| ,\\
		V &= \left( g \hat{A}^\dag + \frac{\Omega}{2} \right) \hat{\sigma}^- ,\\
		\hat{A} &= \left( e^{-i\phi/2}\hat{a}_{\rm CW}+e^{i\phi/2}\hat{a}_{\rm CCW} \right)|L \rangle\langle L| \\
		&\phantom{{}=} + \left( e^{i\phi/2}\hat{a}_{\rm CW}+e^{-i\phi/2}\hat{a}_{\rm CCW} \right) |R \rangle\langle R| \\
		&= \sqrt{2} \left(\cos\frac{\phi}{2} \hat{a}_S + \sin\frac{\phi}{2} \hat{a}_A \hat{\sigma}_{\rm ext}^x \right) .
\end{align*} 
\end{subequations}
Finally, the excited state $|e\rangle$ can be eliminated resulting in the effective Hamiltonian and dissipator~\cite{Reiter2012}
\begin{subequations} \label{eq:eff_nH}
	\begin{align}
		H_{\rm eff, nH} &= H_{g, \rm nH} - V \frac{1}{H_{e, \rm nH}} V^\dag \nonumber \\
		&= H_{g, \rm nH} + \frac{1}{\Delta + i \gamma/2} \left[ g^2 \hat{A}^\dag \hat{A} + \frac{g\Omega}{2} \left( \hat{A}^\dag + \hat{A} \right)  + \frac{\Omega^2}{4} \right] , \\
		\hat{L}_{\rm eff} &= \hat{\sigma}^- \frac{1}{H_{e, \rm nH}} V^\dag = - \frac{g \hat{A} + \frac{\Omega}{2}}{\Delta+i\gamma/2} \left| g\middle\rangle \middle\langle g\right| .
	\end{align}
\end{subequations}
In the dispersive regime, $\langle \hat{A}^\dag\hat{A} \rangle \ll {\rm Re} \langle \hat{A} \rangle $, which allows us to neglect the $\hat{A}^\dag\hat{A}$ term in Eq.~\eqref{eq:eff_nH} and keep the linear terms only. The decay rate of the dissipator $\hat{L}_{\rm eff}$ is on the order of $\gamma (g^2\langle\hat{A}^\dag\hat{A}\rangle+\Omega^2)/\Delta^2$ which, in the dispersive regime, is much smaller than the decay rate of the cavity photon $\kappa$, so $\hat{L}_{\rm eff}$ can be neglected as well as the corresponding imaginary part of $H_{\rm eff, nH}$. Therefore, the original master equation [Eq.~\eqref{eq:me} or \eqref{eq:original_ME}] can be simplified as
\begin{equation} \label{eq:me_eff}
	\frac{\partial}{\partial t} \rho = -i [H_{\rm eff}, \rho] + \kappa \mathcal{D}[\hat{a}_S]\rho + \kappa \mathcal{D}[\hat{a}_A]\rho,
\end{equation}
with 
\begin{equation} \label{eq:Heff}
	\begin{split}
		H_{\rm eff} &= H_S + H_A - J \hat{\sigma}_{\rm ext}^z ,\\		
		H_S &= - \delta \hat{a}_S^\dagger \hat{a}_S + \frac{\Omega_{\rm eff}}{2} \cos\frac{\phi}{2} \left( \hat{a}_S^\dagger + \hat{a}_S \right) ,\\
		H_A &= - \delta \hat{a}_A^\dagger \hat{a}_A + \frac{\Omega_{\rm eff}}{2} \sin\frac{\phi}{2} \left( \hat{a}_A^\dagger + \hat{a}_A \right) \hat{\sigma}_{\rm ext}^{x}.
	\end{split}
\end{equation}
This is the effective Hamiltonian stated in Eq.~\eqref{eq:H_SA}, where the effective Rabi frequency
\begin{equation}
	\Omega_{\rm eff} = \frac{\sqrt{2}g\Omega\Delta}{\Delta^2+\gamma^2/4} \approx \frac{\sqrt{2}g\Omega}{\Delta} .
\end{equation}
Here we note that if one omits the external state, Eqs.~\eqref{eq:me_eff} and \eqref{eq:Heff} can be used to describe the single atom trapped in a single well. Considering the Heisenberg equations of the photon field operators $i\partial_t \hat{a}_{S} = -(\delta + i\kappa/2) \hat{a}_{S} + \frac{\Omega_{\rm eff}}{2} \cos\frac{\phi}{2}$ and $i\partial_t \hat{a}_{A} = -(\delta + i\kappa/2) \hat{a}_{A} + \frac{\Omega_{\rm eff}}{2} \sin\frac{\phi}{2}$, we can obtain the photon numbers of the steady state
\begin{equation}
	n_0 = \left| \alpha_S \right|^2 + \left| \alpha_A \right|^2 = \left| \frac{\Omega_{\rm eff}}{2} \frac{1}{\delta + i\kappa/2 } \right|^2 ,
\end{equation}
with 
\begin{equation}
	\begin{split}
		\alpha_{S} &= \langle \hat{a}_{S} \rangle = \frac{\Omega_{\rm eff}}{2} \frac{1}{\delta+i\kappa/2} \cos\frac{\phi}{2} , \\
		\alpha_{A} &= \langle \hat{a}_{A} \rangle = \frac{\Omega_{\rm eff}}{2} \frac{1}{\delta+i\kappa/2} \sin\frac{\phi}{2} .
	\end{split}
\end{equation}

However, if the atomic motion is included, the $\hat{a}_A$ mode cavity field is different from the ordinary bosonic field, because creating an $A$ mode photon flips the external state. Specifically, the Heisenberg equation of $\hat{a}_A$ becomes $i\partial_t \hat{a}_{A} = -(\delta + i\kappa/2) \hat{a}_{A} + \frac{\Omega_{\rm eff}}{2} \sin\frac{\phi}{2} \hat{\sigma}_{\rm ext}^x$. Now $\hat{\sigma}_{\rm ext}^x$ oscillates at the frequency approximately equal to $2J$, which precludes the quasistatic condition by letting $\partial_t \hat{a}_{A} = 0$. To resolve this problem, we notice that the transitions of $|n_A,+\rangle \to |(n-1)_A,-\rangle$ and $|n_A,-\rangle \to |(n-1)_A,+\rangle$ can be triggered by $\hat{a}_A \hat{\sigma}_{\rm ext}^-$ and $\hat{a}_A \hat{\sigma}_{\rm ext}^+$, respectively. Therefore, we consider the Heisenberg equations of $\hat{a}_S$, $\hat{a}_A \hat{\sigma}_{\rm ext}^+$ and $\hat{a}_A \hat{\sigma}_{\rm ext}^-$, which read
\begin{equation}  \label{eq:Heinsenberg_quasistatic} 
	\begin{split}
		i\frac{\partial}{\partial t} \hat{a}_S &= 
			- \left(\delta + i\frac{\kappa}{2}\right) \hat{a}_S 
			+ \frac{\Omega_{\rm eff}}{2} \cos\frac{\phi}{2} , \\
		i\frac{\partial}{\partial t} \hat{a}_A \hat{\sigma}_{\rm ext}^+ &= 
			- \left(\delta + 2J + i\frac{\kappa}{2}\right) \hat{a}_A \hat{\sigma}_{\rm ext}^+ \\
			&\phantom{{}=} + \frac{\Omega_{\rm eff}}{2} \sin\frac{\phi}{2} \left[ \hat{\sigma}_{\rm ext}^+\hat{\sigma}_{\rm ext}^- - \left( \hat{a}_A + \hat{a}_A^\dag \right)\hat{a}_A \hat{\sigma}_{\rm ext}^z  \right] , \\
		i\frac{\partial}{\partial t} \hat{a}_A \hat{\sigma}_{\rm ext}^- &= 
			- \left(\delta - 2J + i\frac{\kappa}{2}\right) \hat{a}_A \hat{\sigma}_{\rm ext}^- \\
			&\phantom{{}=} + \frac{\Omega_{\rm eff}}{2} \sin\frac{\phi}{2} \left[ \hat{\sigma}_{\rm ext}^-\hat{\sigma}_{\rm ext}^+ + \left( \hat{a}_A + \hat{a}_A^\dag \right)\hat{a}_A \hat{\sigma}_{\rm ext}^z  \right] .
	\end{split}
\end{equation}
Then we can let the left-hand side equal zero for the quasistatic solutions for the given external state, because the evolution rate of $\hat{\sigma}_{\rm ext}^z$, as well as $\hat{\sigma}_{\rm ext}^\pm\hat{\sigma}_{\rm ext}^\mp$, is on the order of $\Gamma \ll J$ (discussed below). In the dispersive regime, the photon number is small, and we can further neglect the normal order terms with multiple $\hat{a}_A$ terms on the right-hand side. Therefore, the adiabatic elimination can be carried out by taking the quantum expectation values of Eq.~\eqref{eq:Heinsenberg_quasistatic}, which yields 
\begin{subequations} \label{eq:adiabatic_operators}
	\begin{align}  
			\left\langle \hat{a}_S \right\rangle &\approx
				\frac{\Omega_{\rm eff}/2}{\delta+i\kappa/2} \cos\frac{\phi}{2} ,\\
			\left\langle \hat{a}_A\hat{\sigma}_{\rm ext}^+ \right\rangle &\approx 
				\frac{\Omega_{\rm eff}/2}{\delta+2J+i\kappa/2} \sin\frac{\phi}{2} \left\langle \hat{\sigma}_{\rm ext}^+\hat{\sigma}_{\rm ext}^- \right\rangle ,\\
			\left\langle \hat{a}_A\hat{\sigma}_{\rm ext}^- \right\rangle &\approx 
				\frac{\Omega_{\rm eff}/2}{\delta-2J+i\kappa/2} \sin\frac{\phi}{2} \left\langle \hat{\sigma}_{\rm ext}^-\hat{\sigma}_{\rm ext}^+ \right\rangle .
	\end{align}
\end{subequations}
We introduce the reduced density matrix of the external state, $\rho_{\rm ext} = {\rm Tr}_{\rm ph}(\rho)$, and $\rho_{\rm ext}^{++} = \left\langle \hat{\sigma}_{\rm ext}^-\hat{\sigma}_{\rm ext}^+ \right\rangle$, $\rho_{\rm ext}^{--} = \left\langle \hat{\sigma}_{\rm ext}^+\hat{\sigma}_{\rm ext}^- \right\rangle$, $\rho_{\rm ext}^{+-} = \left\langle \hat{\sigma}_{\rm ext}^+ \right\rangle$, and $\rho_{\rm ext}^{-+} = \left\langle \hat{\sigma}_{\rm ext}^- \right\rangle$. Based on Eq.~\eqref{eq:adiabatic_operators}, we can use the substitution
\begin{equation} \label{eq:op_sub}
	\begin{split}
		\hat{a}_S &\to \frac{\Omega_{\rm eff}/2}{\delta+i\kappa/2} \cos\frac{\phi}{2} ,\\
		\hat{a}_A &\to \frac{\Omega_{\rm eff}/2}{\delta+2J+i\kappa/2} \sin\frac{\phi}{2} \overleftarrow{\hat{\sigma}}_{\rm ext}^- \\
		&\phantom{{}\to} + \frac{\Omega_{\rm eff}/2}{\delta-2J+i\kappa/2} \sin\frac{\phi}{2} \overleftarrow{\hat{\sigma}}_{\rm ext}^+ ,
	\end{split}
\end{equation}
where the notation $\overleftarrow{\sigma}$ emphasizes that when substituting Eq.~\eqref{eq:adiabatic_operators}, the operator $\hat{a}_A$ must be placed on the left side of $\hat{\sigma}$. With this substitution and using the identity $\hat{\sigma}_{\rm ext}^- \hat{\sigma}_{\rm ext}^+ + \hat{\sigma}_{\rm ext}^+ \hat{\sigma}_{\rm ext}^- = \mathbb{I}$, the complex light fields of the $\hat{a}_S$ and $\hat{a}_A$ modes are given by [the results of Eq.~\eqref{eq:light_class}]
\begin{equation} \label{eq:mf_alpha}
	\begin{split}
		\alpha_S &= \frac{\Omega_{\rm eff}/2}{\delta+i\kappa/2} \cos\frac{\phi}{2} ,\\
		\alpha_A &= \left\langle \hat{a}_A \left(\hat{\sigma}_{\rm ext}^- \hat{\sigma}_{\rm ext}^+ + \hat{\sigma}_{\rm ext}^+ \hat{\sigma}_{\rm ext}^- \right) \right\rangle \\
		&= 
			\frac{\Omega_{\rm eff}/2}{\delta+2J+i\kappa/2} \sin\frac{\phi}{2} \left\langle \hat{\sigma}_{\rm ext}^- \right\rangle  \\
		&\phantom{{}=} + \frac{\Omega_{\rm eff}/2}{\delta-2J+i\kappa/2} \sin\frac{\phi}{2} \left\langle \hat{\sigma}_{\rm ext}^+ \right\rangle .
	\end{split}
\end{equation} 
The photon numbers yield
\begin{widetext}
\begin{equation} \label{eq:mf_n}
	\begin{split}
		n_S &= \left| \frac{\Omega_{\rm eff}/2}{\delta+i\kappa/2} \cos\frac{\phi}{2} \right|^2 ,\\
		n_A &= \left\langle \hat{a}_A^\dag \left(\hat{\sigma}_{\rm ext}^+ \hat{\sigma}_{\rm ext}^- + \hat{\sigma}_{\rm ext}^- \hat{\sigma}_{\rm ext}^+ \right) \hat{a}_A \right\rangle \\
		&=  \left| \frac{\Omega_{\rm eff}/2}{\delta+2J+i\kappa/2} \sin\frac{\phi}{2} \right|^2 \left\langle \hat{\sigma}_{\rm ext}^+ \hat{\sigma}_{\rm ext}^- \right\rangle 
		 + \left| \frac{\Omega_{\rm eff}/2}{\delta-2J+i\kappa/2} \sin\frac{\phi}{2} \right|^2 \left\langle \hat{\sigma}_{\rm ext}^- \hat{\sigma}_{\rm ext}^+ \right\rangle .
	\end{split}
\end{equation} 
Therefore, we obtain the photon numbers of the CW and CCW modes, which read
\begin{equation} \label{eq:photon_solution}
	\begin{split}
		n_{\rm CW/CCW} &= \frac{1}{2} \left\langle \left( \hat{a}_S \pm i \hat{a}_A \right)^\dag \left( \hat{a}_S \pm i \hat{a}_A \right) \right\rangle 
		= \frac{1}{2} \left[ n_S + n_A \pm i \left( \alpha_S^* \alpha_A - \alpha_A^* \alpha_S \right) \right] \\
		&= \frac{n_0}{2} \left[
			\cos^2\frac{\phi}{2} 
			+ \sin^2\frac{\phi}{2} \left(
				\left| \frac{\delta+i\kappa/2}{\delta+2J+i\kappa/2} \right|^2 \rho_{\rm ext}^{--} +
				\left| \frac{\delta+i\kappa/2}{\delta-2J+i\kappa/2} \right|^2 \rho_{\rm ext}^{++} \right) 
			\pm {\rm Im} \frac{ 4 J \delta \sin\phi }{ (\delta+2J+i\kappa/2)(\delta-2J-i\kappa/2) } \rho_{\rm ext}^{-+} \right].
	\end{split}
\end{equation}
\end{widetext}

\section{Dynamics of the atomic motion influenced by the cavity photons}
\label{appx:dyn}

In order to obtain the effective master equation of the atomic motion, we eliminate the photonic excited states with the adiabatic approximation. According to Eq.~\eqref{eq:H_SA} and \eqref{eq:Heff}, the $\hat{a}_S$ mode photons are decoupled from the atomic motion and thereby can be traced out directly. In the dispersive regime, the $\hat{a}_A$ mode photon number can be truncated at $n_A \leqslant 1$, which results in four states $|n_A, m\rangle \in \{ |0_A, +\rangle, |0_A, -\rangle, |1_A, +\rangle, |1_A, -\rangle \}$. The non-Hermitian Hamiltonian reads
\begin{equation}
	H_{\rm nH} = \begin{pmatrix}
		-J & & & \frac{\Omega_{\rm eff}}{2} \sin\frac{\phi}{2} \\
		&  J & \frac{\Omega_{\rm eff}}{2} \sin\frac{\phi}{2} & \\
		& \frac{\Omega_{\rm eff}}{2} \sin\frac{\phi}{2} & -J-\delta-i\frac{\kappa}{2} \\
		\frac{\Omega_{\rm eff}}{2} \sin\frac{\phi}{2} & & & J-\delta-i\frac{\kappa}{2}
	\end{pmatrix} .
\end{equation}
The quantum jump operators include $\hat{L}_- = \left|0_A, +\middle\rangle \middle\langle 1_A, -\right|$ and $\hat{L}_+ = \left|0_A, -\middle\rangle \middle\langle 1_A, +\right|$ with rates both equal to $\kappa$. To perform the adiabatic elimination, we introduce the projection operator $\mathcal{P} = \left|0_A, +\middle\rangle \middle\langle0_A, +\right| + \left|0_A, -\middle\rangle \middle\langle0_A, -\right|$ and $\mathcal{Q} = \left|1_A, +\middle\rangle \middle\langle1_A, +\right| + \left|1_A, -\middle\rangle \middle\langle1_A, -\right|$. Noticing that $\mathcal{P} H_{\rm nH} \mathcal{P}$ is already diagonalized, we define the interaction operators acting on each of the vacuum states, $V_l = \left|l\middle\rangle \middle\langle l\right| H_{\rm nH} \mathcal{Q}$, with $|l\rangle \in \{ |0_A, +\rangle, |0_A, -\rangle \}$~\cite{Reiter2012}. Then we obtain the effective non-Hermitian Hamiltonian
\begin{equation}
	\begin{split}
		H_{\rm ext, nH} &= \mathcal{P} H_{\rm nH} \mathcal{P} - V \sum_{l} \frac{1}{\mathcal{Q}H_{\rm nH}\mathcal{Q} - E_l} V_l^\dag \\
			&= - \left( J - \frac{\Omega_{\rm eff}^2}{4} \sin^2\frac{\phi}{2} \frac{1}{\delta-2J+i\kappa/2} \right) \left|0_A, +\middle\rangle \middle\langle0_A, +\right| \\
			&\phantom{{}=} + 
				\left( J + \frac{\Omega_{\rm eff}^2}{4} \sin^2\frac{\phi}{2} \frac{1}{\delta+2J+i\kappa/2} \right) \left|0_A, -\middle\rangle \middle\langle0_A, -\right| ,
	\end{split}
\end{equation}
and the dissipation operators
\begin{equation}
	\begin{split}
		\hat{L}_{\pm,\rm eff} &= \left|0_A, \mp\middle\rangle\middle\langle 1_A, \mp\right| \sum_l \frac{1}{\mathcal{Q}H_{\rm nH}\mathcal{Q} - E_l} V_l^\dag \\
			&= -\frac{\Omega_{\rm eff}}{2} \sin\frac{\phi}{2} \frac{1}{\delta \mp 2J+i\kappa/2} \left|0_A, \mp \middle\rangle \middle\langle0_A, \pm\right| .
	\end{split}
\end{equation}
Therefore, we obtain the effective master equation of the atomic motion, which reads
\begin{equation}
	\begin{split}
		\frac{\partial}{\partial t} \rho_{\rm ext} &=
		-i \left( H_{\rm ext, nH} \rho_{\rm ext} - \rho_{\rm ext} H_{\rm ext, nH}^\dag \right) \\
		&\phantom{{}=} 
		+ \kappa  \mathcal{D}[\hat{L}_{+,\rm eff}]\rho_{\rm ext} + \kappa \mathcal{D}[\hat{L}_{-,\rm eff}]\rho_{\rm ext} .
	\end{split}
\end{equation}
Then it can be simplified to Eqs.~\eqref{eq:me_ext} and \eqref{eq:shift}, 
\begin{equation} \label{eq:me_motion}
	\begin{split}
		\frac{\partial}{\partial t} \rho_{\rm ext} &=
			-i [-J' \hat{\sigma}_{\rm ext}^z + J_+ \left| + \middle\rangle\middle\langle + \right| + J_- \left| - \middle\rangle\middle\langle - \right|, \rho_{\rm ext}] \\
			&\phantom{{}=}
			+ \Gamma_+ \mathcal{D}[\hat{\sigma}_{\rm ext}^+]\rho_{\rm ext}
			+ \Gamma_- \mathcal{D}[\hat{\sigma}_{\rm ext}^-]\rho_{\rm ext} .
	\end{split}
\end{equation}
with $J_\pm$ and $\Gamma_\pm$ originated from the ac Stark effect, which read
\begin{equation}
	J_\pm - i \frac{\Gamma_\pm}{2} = \frac{\Omega_{\rm eff}^2}{4} \frac{1}{\delta \mp 2J + i \kappa/2} \sin^2\frac{\phi}{2} .
\end{equation} 

The decoherence rates $\Gamma_\pm$ correspond to the incoherent population transfer from the subspace $\mathscr{H}_\pm$ to the other. The total decoherence rate which also acts on the atomic motion and cavity field reads
\begin{equation} \label{eq:Gamma_tot}
	\begin{split}
		\Gamma &= \Gamma_+ + \Gamma_- \\
		&= \kappa \frac{\Omega_{\rm eff}^2}{4} \sin^2\frac{\phi}{2} \left(  \frac{\sin^2\frac{\phi}{2}}{(\delta-2J)^2+\kappa^2/4} + \frac{\sin^2\frac{\phi}{2}}{(\delta+2J)^2+\kappa^2/4} \right).
	\end{split}
\end{equation}

\begin{figure}
	\centering
	\includegraphics[width=\linewidth]{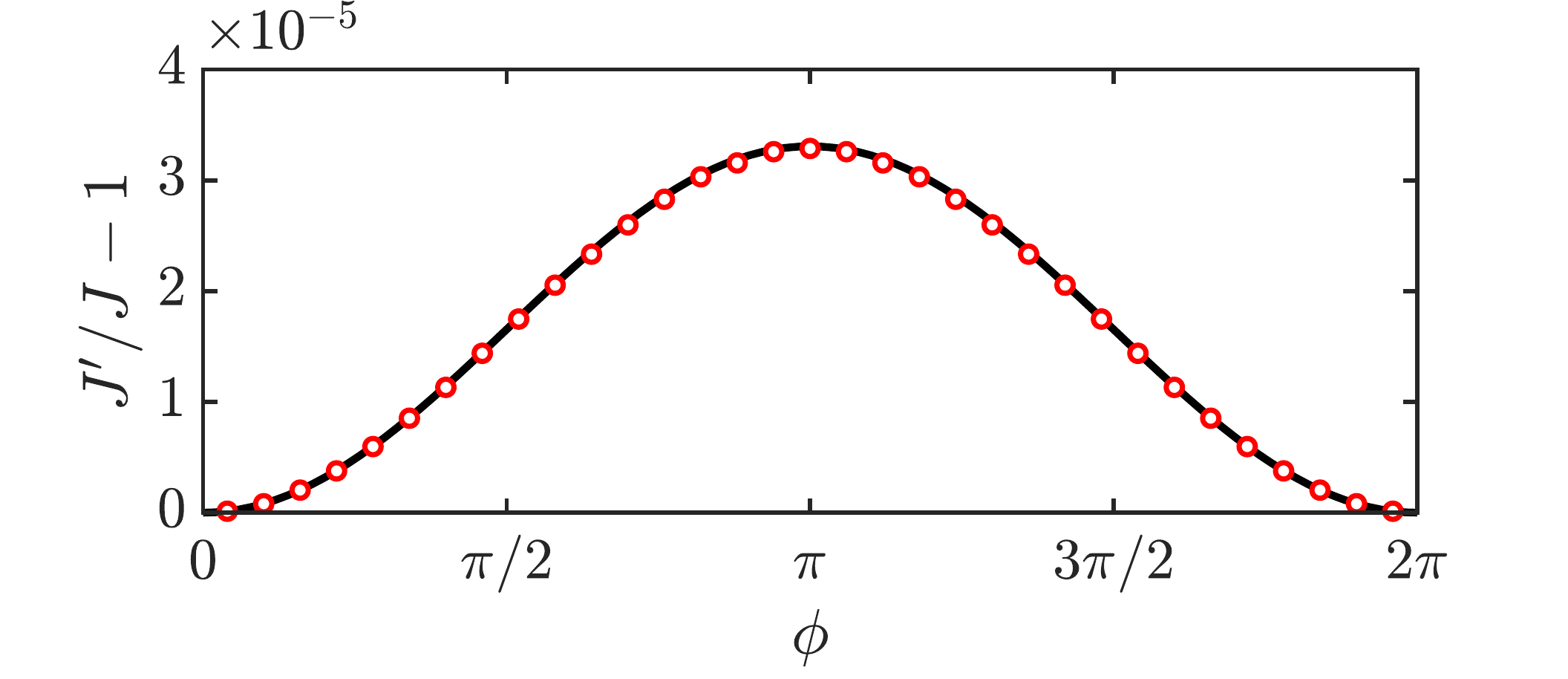}
	\caption{ \label{fig:Rabi_freq} Relative correction of the tunneling rate due to the optical potential. The black line is the analytical results according to Eq.~\eqref{eq:me_motion} and the red circles are numerical results obtained from Fig.~\ref{fig:chirality} with different $\phi$ and fitted sinusoidally in the long-time regime ($Jt>4500$). }
\end{figure}

The tunneling rate is modified by
\begin{equation}
	\begin{split}
		J' &= J - \frac{J_+ - J_-}{2} \\
		&= J + \frac{1}{2} \left( \frac{\Omega_{\rm eff}}{2} \sin\frac{\phi}{2} \right)^2 {\rm Re}\left[ \frac{-4J}{(\delta+i\kappa/2)^2-4J^2}\right] .		
	\end{split}
\end{equation}
As shown in Fig.~\ref{fig:Rabi_freq}, we compare $J'$ and $J$ with the same parameters as used in Fig.~\ref{fig:chirality} and find that the relative correction of the tunneling rate is on the order of $10^{-5}$. Therefore, we can neglect the correction to the tunneling rate for the steady-state solutions. However, in the long-term evolution when $\Gamma t \gg 1$, the optical potential will lead to a non-negligible phase shift due to the modification of the tunneling frequency.

Then, Eq.~\eqref{eq:me_motion} produces the semiclassical equation of the atomic motion
\begin{equation}
	\frac{\partial}{\partial t} \begin{pmatrix}
		\left\langle \hat{\sigma}^x \right\rangle \\ \left\langle \hat{\sigma}^y \right\rangle \\ \left\langle \hat{\sigma}^z \right\rangle
	\end{pmatrix} = \begin{pmatrix}
		-\Gamma/2 & -2J' &  \\
		2J' & -\Gamma/2  &  \\
		 &  & -\Gamma  
	\end{pmatrix} \begin{pmatrix}
		\left\langle \hat{\sigma}^x \right\rangle \\ \left\langle \hat{\sigma}^y \right\rangle \\ \left\langle \hat{\sigma}^z \right\rangle
	\end{pmatrix} + \begin{pmatrix}
		0 \\ 0 \\ \Gamma_- - \Gamma_+
	\end{pmatrix} ,
\end{equation}

\begin{figure}
	\centering
	\includegraphics[width=\linewidth]{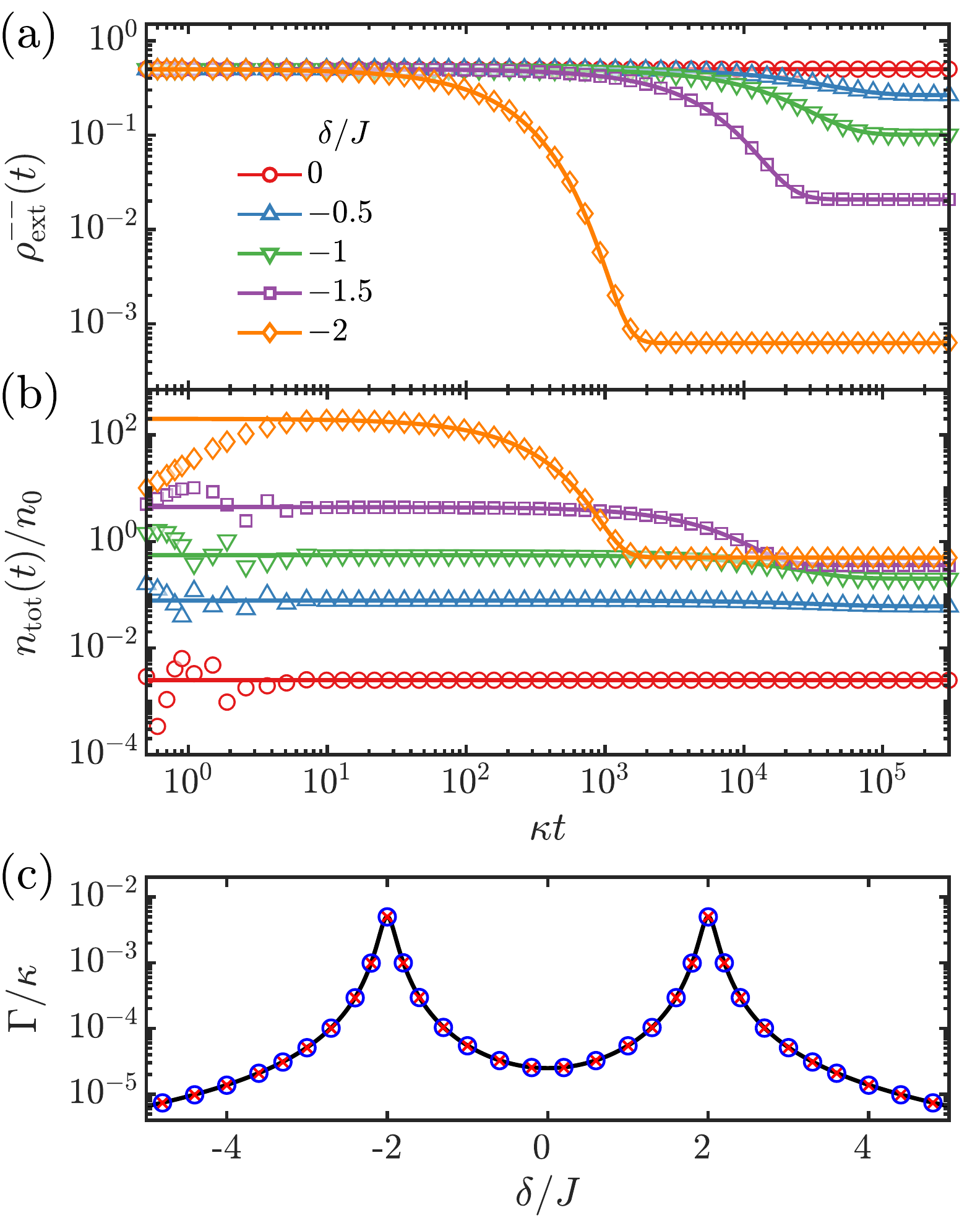}
	\caption{ \label{fig:dynamics} Evolution of (a) the atomic population on the $|-\rangle = (|L\rangle - |R\rangle)/\sqrt{2}$ state and (b) the total cavity photon numbers $n_{\rm tot}(t)$. The double-well spacing $\phi=\pi/4$ ($d=\lambda/2$) and the initial state is $|0,0,L,g\rangle$. The scatters are numerical simulations based on the master equation for different $\delta$ and the solid lines are analytical solutions according to Eq.~\eqref{eq:motional_solution} for the atomic external state and Eq.~\eqref{eq:photon_solution} for the photon numbers. (c) Total relaxation rate $\Gamma = \Gamma_+ + \Gamma_-$ in the long-time regime. The black line is the analytical solution \eqref{eq:Gamma_tot} and the points are exponential fitting results of the numerical simulations of atomic motion $\rho_{\rm ext}(t)$ (red crosses), and photon numbers $n_{\rm tot}(t)$ (blue circles). }
\end{figure}

Letting the initial state be $|L\rangle$, with $\left\langle \hat{\sigma}^x (0) \right\rangle=1$ and $\left\langle \hat{\sigma}^y (0) \right\rangle=\left\langle \hat{\sigma}^z (0) \right\rangle=0$, the solution reads
\begin{equation} \label{eq:motional_solution}
	\begin{split}
		\left\langle \hat{\sigma}^x (t) \right\rangle &= e^{-(\Gamma/2)t} \cos 2J't ,\\
		\left\langle \hat{\sigma}^y (t) \right\rangle &= e^{-(\Gamma/2)t} \sin 2J't ,\\
		\left\langle \hat{\sigma}^z (t) \right\rangle &= \frac{\Gamma_- - \Gamma_+}{\Gamma} \left( 1 - e^{-\Gamma t}\right) .
	\end{split}
\end{equation}
Substituting Eq.~\eqref{eq:motional_solution} into Eq.~\eqref{eq:photon_solution}, we obtain the analytical solutions of the instantaneous photon numbers. For the steady state, $\left\langle \hat{\sigma}^x (\infty) \right\rangle=\left\langle \hat{\sigma}^y (\infty) \right\rangle=0$ and 
\begin{equation}
	\begin{split}
		\left\langle \hat{\sigma}^z (\infty) \right\rangle &= \frac{\Gamma_- - \Gamma_+}{\Gamma} = \frac{-4\delta J'}{\delta^2 + 4J'^2 + \kappa^2/4}, \\
		n_{\rm tot} &= n_0 \left( \cos^2\frac{\phi}{2} + \frac{1}{1+4J'^2/(\delta^2+\kappa^2/4)} \sin^2\frac{\phi}{2} \right) .
	\end{split}
\end{equation}
With the approximation $J' \approx J$, we obtain the steady-state solution of Eq.~\eqref{eq:steady}.

In Fig.~\ref{fig:dynamics}(a) and \ref{fig:dynamics}(b) we present a comparison between the analytical solutions (solid lines) and numerical simulations (points) of the dynamics. The double-well spacing for the simulation is $\phi=\pi/4$ ($d=\lambda/2$), the atom is initially in the $|L\rangle$ state, and the cavity is set to vacuum. We evaluate the saturation dynamics of the atomic tunneling using $\rho_{\rm ext}^{--}(t)$ and compare it with Eq.~\eqref{eq:motional_solution}. Similarly, we compare the photon numbers $n_{\rm tot}(t)$ with the adiabatic solution given in Eq.~\eqref{eq:photon_solution}. The numerical and analytical solutions initially do not match during the short-time dynamics of the internal states when $\kappa t < 10$. However, after $\kappa t > 10$, the analytical solutions can accurately describe the numerical simulations. We observe that for $\delta < -2J$, as $J$ increases, the relaxation of the atomic motion slows down. Moreover, the photon numbers experience longer quasistatic plateaus before equilibrium.

In Fig.~\ref{fig:dynamics}(c), we have fitted $\rho_{\rm ext}(t)$ and $n_{\rm tot}(t)$ with an exponential function to obtain their relaxation rates compared to Eq.~\eqref{eq:motional_solution}. We observe that the numerical results agree well with the analytical solution, and the relaxation rates of the cavity photons and atomic motion are similar, implying that the relaxation of the photons keeps pace with the atomic motion. When $\delta = \pm 2J$, $\Gamma$ reaches the maximum value of $(g\Omega/\Delta)^2 / 2\kappa \sin^2\frac{\phi}{2}$ due to the resonant coupling between $|0_A, \pm\rangle$ and $|1_A, \mp\rangle$ states. This process leads to rapid relaxation, similar to the sideband cooling mechanism observed in optomechanics and ion traps. If $\delta \neq \pm 2J$, we expect the atom to tunnel in the double well without significant decoherence, and the light fields to adiabatically follow the atomic motion with oscillating directionality.

\section{Photon correlation of the steady state}
\label{appx:corr}

As discussed in the main text, one can separate the Hilbert space into two subspaces based on their parity. For the steady state, the density matrix can be written as (under the basis of $|n_S, n_A, m\rangle$)
\begin{align}
	\rho_{\rm ss} &= P_+ |\psi_+\rangle \langle\psi_+| + P_- |\psi_-\rangle \langle\psi_-| ,\\
	|\psi_+\rangle &\approx c_{00+}|00+\rangle + c_{10+}|10+\rangle + c_{01-}|01-\rangle  \nonumber \\
	&\phantom{{}=} + c_{20+}|20+\rangle + c_{11-}|11-\rangle + c_{02+}|02+\rangle , \nonumber \\
	|\psi_-\rangle &\approx c_{00-}|00-\rangle + c_{10-}|10-\rangle + c_{01+}|01+\rangle \nonumber \\
	&\phantom{{}=} + c_{20-}|20-\rangle + c_{11+}|11+\rangle + c_{02-}|02-\rangle . \nonumber 
\end{align}
In the dispersive regime, the coefficients can be derived perturbatively, which read
\begin{align*}
		&\qquad P_\pm = \frac{1}{2} \left( 1 \mp \frac{4 \delta J}{\delta^2+4J^2+\kappa^2/4} \right)  ,\\
		&c_{00+} = c_{00-} \approx 1 ,\qquad
		c_{10+} = c_{10-} \approx \frac{\Omega_{\rm eff}}{2} \frac{\cos\frac{\phi}{2}}{\delta+i\kappa/2} ,\\
		&c_{01-} \approx \frac{\Omega_{\rm eff}}{2} \frac{\sin\frac{\phi}{2}}{\delta-2J+i\kappa/2} ,\qquad
		c_{01+} \approx \frac{\Omega_{\rm eff}}{2} \frac{\sin\frac{\phi}{2}}{\delta+2J+i\kappa/2} ,\\
		&c_{11-} \approx c_{10+} c_{01-} ,\qquad 
		c_{11+} \approx c_{10-} c_{01+} ,\\
		&c_{20+} = c_{20-} \approx \sqrt{2} \frac{\Omega_{\rm eff}}{2} \frac{\cos\frac{\phi}{2}}{2\delta+i\kappa}  c_{10+}  ,\\
		&c_{02+} \approx \sqrt{2} \frac{\Omega_{\rm eff}}{2} \frac{\sin\frac{\phi}{2}}{2\delta+i\kappa}  c_{01-} ,\qquad
		c_{02-} \approx \sqrt{2} \frac{\Omega_{\rm eff}}{2} \frac{\sin\frac{\phi}{2}}{2\delta+i\kappa}  c_{01+} .
\end{align*}
The steady-state probabilities, $P_\pm$, are derived in Appendix~\ref{appx:dyn}.
For the photon mode $\hat{a}$, specifically, $\hat{a}_{\rm CW} = (\hat{a}_S + i \hat{a}_A)/\sqrt{2}$ and $\hat{a}_{\rm CCW} = (\hat{a}_S - i \hat{a}_A)/\sqrt{2}$, the second-order correlation is 
\begin{equation}
	g^{(2)}(\tau) = \frac{{\rm Tr} \left[ \hat{a} U(\tau) \hat{a} \rho_{\rm ss} \hat{a}^\dag U^\dag(\tau) \hat{a}^\dag \right]  }{ {\rm Tr} \left( \hat{a}^\dag \hat{a} \rho_{\rm ss} \right)^2 }. 
\end{equation}
The evolution propagator represented in the manifold consists of vacuum states ($\mathbb{H}_0=\{ |00+\rangle, |00-\rangle \}$) and single-photon states ($\mathbb{H}_1=\{ |10+\rangle, |10-\rangle, |01+\rangle, |01-\rangle \}$), which reads
\begin{equation}
	\begin{split}
		U(\tau) &\approx \sum_{j \in \mathbb{H}_0 \cup \mathbb{H}_1 } e^{-i \varepsilon_j t} \left| j \middle\rangle\middle\langle j \right| \\
		&\phantom{{}=} - \sum_{\substack{j \in \mathbb{H}_1 \\ k \in \mathbb{H}_0}} v_{jk} \frac{e^{-i \varepsilon_j t} - e^{-i \varepsilon_k t}}{\varepsilon_j - \varepsilon_k}  \left| j \middle\rangle\middle\langle k \right| + {\rm H.c.}.
	\end{split}
\end{equation}
The non-zero coefficients are
\begin{align*}
	&\varepsilon_{00\pm} = \mp J\ , \quad
	 \varepsilon_{10\pm} = \varepsilon_{01\pm} = \delta \mp J + i\kappa/2\ , \\
	&v_{10\pm,00\pm} = \frac{\Omega_{\rm eff}}{2} \cos\frac{\phi}{2} \ , \quad 
	v_{01\mp,00\pm} = \frac{\Omega_{\rm eff}}{2} \sin\frac{\phi}{2} \ .
\end{align*}

The correlation functions of the CW and CCW modes are the same, and both yield
\begin{widetext}
\begin{equation}
	\begin{split}
		g^{(2)}(\tau) &=  \left( 1 + \frac{4J^2}{\delta^2+\kappa^2/4} \right) \left[ 1 + \left( 1 + \frac{4J^2}{\delta^2+\kappa^2/4} \right) \cot^2\frac{\phi}{2} \right]^{-2}  \left( \frac{1}{2} \left| \frac{(\delta+i\kappa/2)e^{-2iJ\tau}-2Je^{-i\delta\tau-\kappa\tau/2}}{\delta+2J+i\kappa/2} - \frac{\delta-2J+i\kappa/2}{\delta+i\kappa/2} \cot^2\frac{\phi}{2} \right|^2 + \right. \\
		&\phantom{{}=} \qquad \left. \frac{1}{2} \left| \frac{(\delta+i\kappa/2)e^{2iJ\tau}+2Je^{-i\delta\tau-\kappa\tau/2}}{\delta-2J+i\kappa/2} - \frac{\delta+2J+i\kappa/2}{\delta+i\kappa/2} \cot^2\frac{\phi}{2} \right|^2 + 4 \cot^2\frac{\phi}{2} \cos^2J\tau \right).
	\end{split}
\end{equation}
For a simple case, let $\delta=0$. Then the correlation function reads [Eq.~\eqref{eq:g2}]
\begin{equation}
	g^{(2)}(\tau) = 1 + \left[ 1 + \left( 1 + \frac{4J^2}{\kappa^2/4} \right) \cot^2\frac{\phi}{2} \right]^{-2} \left( \frac{4J^2}{\kappa^2/4} e^{-\kappa\tau} + \frac{4J}{\kappa/2} e^{-\kappa\tau/2} \sin2J\tau \right).
\end{equation}
\end{widetext}

%

\end{document}